\documentclass[%
showkeys,
amsmath,
amssymb,
aps,
superscriptaddress,
twocolumn,
]
{revtex4-1}

\usepackage[normalem]{ulem}

\usepackage[dvips]{graphicx}
\usepackage{amssymb,amsfonts,amsmath}
\usepackage[sort&compress]{natbib}
\usepackage[version=3]{mhchem} 
\usepackage{color}
\usepackage[outdir=./]{epstopdf}
\usepackage{dcolumn}
\usepackage{bm}
\usepackage{booktabs,array}
\usepackage[referable]{threeparttablex}
\usepackage{tabularx}
\usepackage{multirow}
\usepackage{color}
\usepackage[english]{babel}

\begin{document}


\title{Neural network learns physical rules for \\copolymer translocation through amphiphilic barriers}


\author{Marco Werner}
\affiliation {Leibniz-Institut f\"{u}r Polymerforschung Dresden e.V., Hohe Stra\ss e 6, 01069 Dresden, Germany}

\author{Yachong Guo}
\affiliation {National Laboratory of Solid State Microstructure, Department of Physics, Nanjing University, Nanjing 210093, China}
	
\author{Vladimir A. Baulin$^*$}
\affiliation {Departament d'Enginyeria Qu\'{\i}mica, Universitat Rovira i Virgili 26 Av.
	dels Paisos Catalans, 43007 Tarragona Spain}
\date{\today }


\begin{abstract}
Recent development in computer processing power leads to new paradigms of how problems in many-body physics and especially polymer physics can be addressed. GPU parallel processors can be employed to generate millions of independent configurations of polymeric molecules of heterogeneous sequence in complex environments at a second, and concomitant free-energy landscapes estimated. Resulting data bases that are complete in terms of polymer sequence and architecture are a powerful training basis for multi-layer artificial neural networks, whose internal representations will potentially lead to a new physical viewpoint in how sequence patterns are linked to effective polymer properties and response to the environment. In our example, we consider the translocation time of a copolymer through an amphiphilic bilayer membranes as a function of binary sequence of hydrophilic and hydrophobic units. First we demonstrate that massively parallel Rosenbluth sampling for all possible sequences of a polymer allows for meaningful dynamic interpretation in terms of the mean first escape times through the membrane. Second we train a multi-layer perceptron, and show by a systematic reduction of the training set to a narrow window of translocation times, that the neural network develops internal representations of the physical rules mapping sequence to translocation times. In particular, based on the narrow training set, the network predicts the correct order of magnitude of translocation times in a window that is more than 8 orders of magnitude wider than the training window.
\end{abstract}

\maketitle

\section{Introduction}\label{sec:level1}

Polymers are many-body physical objects; in order to describe their equilibrium state and dynamics, one needs to map a one-dimensional connectivity rule into a free-energy landscapes in three-dimensional space. Rigorous theoretical descriptions typically capture only simple boundary cases such as homopolymers or copolymers with periodic structure, as they follow bottom-up approaches starting with the local interactions on the monomer level, or consider the self-similarity of self-avoiding walks on the largest scales. The sequence space available by current polymer chemistry \cite{lutzDefining2017,lutzSequenceControlled2013,rahmanMacromolecularclustered2018} or in biopolymers exceeds the limits for closed physical descriptions. By massively parallel conformation sampling for the full binary sequence space of a copolymer we show in this work that the intricacies of polymer sequence-property relationships can be subtle and unexpected already when considering relatively simple inhomogeneities of the environment on the scale of the polymer size.


The laws of physics are, however, normally simple by means of requiring a relatively small number of parameters as compared to machine learning (ML) algorithms such as artificial neural networks (NN). Latter can approximate any function \cite{cybenkoApproximation1989} given that at least one hidden layer of neurons with sigmoid activation functions exists. At a first sight, their generalization performance seems to rely on the property of a universal fitting black box that when fed with unseen input data,  will interpolate the result being thus capable to map even high-dimensional energy surfaces \cite{behlerGeneralized2007}. Equivalent fitting challenges as latter example can be satisfied well also with other ML approaches such as combinations of Gaussian kernels~\cite{bartokGaussian2010}. The stacking of layer to multi-layer non-linear filters, however, seems to mark a qualitative landmark as compared to shallow ML algorithms in such that they develop internal representations of the input information that correspond to a hierarchy of abstraction levels. The distinguished abstraction performance makes so-called deep neural nets (DNN) particularly efficient when confronted with multiple tasks simultaneously, for instance, in finding quantitative structure-property or -activity relationships (QSPR/QSAR)~\cite{caruanaMultitask1997,dahlMultitask2014,maDeep2015,hughesModeling2016,ramsundarMassively2015}. Recent advances in exploiting NN for physical problems show, that NNs can be employed for determining the essential order parameters necessary for predicting a state in future~\cite{itenDiscovering2018}, or classifying a magnetic phases~\cite{carrasquillaMachine2017}.

The mentioned problems of sequence-property mapping in polymers physics are perfect challenges for the exploration by ML methods in order to find hidden abstraction levels and potentially extract new semantic information on theoretical level. The modern stage of computer science has the potential to encourage qualitative jumps in understanding polymer physics fed from three lines of recent development: The acceleration of conformation sampling by highly parallel processors and especially graphics processing units (GPU), the development of advanced sampling techniques for free energy calculation like~\cite{wangEfficient2001,renKinetics2018,guoGPU2017}, and the recent algorithmic advances observed in machine learning~\cite{itenDiscovering2018,carrasquillaMachine2017}. 

Simplified boundary cases, are accessible for analytic theory where polymer backbones are sufficiently homogeneous or provide regular patterns in case of copolymers. For instance, the translocation time of polymer chains through a nano-pore on the scale of one monomer has been first described theoretically for homogeneous backbones~\cite{muthukumarPolymer1999,muthukumarTranslocation2001} expressed in terms of scaling relations, and later on extended to block copolymers~\cite{muthukumarTheory2002}.
In case of pores of the size of a single monomer, one may write the general solution of the backward Kolmogorov equation for general polymer sequences characterized by the monomer's chemical potential inside the pore~\cite{muthukumarTheory2002}. In case of heterogeneously charged polymers dragged by an external field through nano-pores, however, an unexpected strong sensitivity to electric field is subject of further investigation where the effective forces acting on the uncharged monomers due to connectivity still await rigorous theoretical attention~\cite{mirigianTranslocation2012}. As soon as local conformation entropy of the polymer comes into play by widening the pore to a finite diameter and length~\cite{wongPolymer2008,sunTrapped2019} a general expression as a function of sequence seems unreachable in the moment for both charged and uncharged polymers. The outstanding of analytic theoretical mappings between sequence and translocation time meanwhile does not exclude technical applications of nano-pore translocation for DNA sequencing~\cite{kasianowiczCharacterization1996,liDNA2003,clarkeContinuous2009}.

The picture is similar when considering the translocation of a polymer through a lipid membrane by direct penetration of the membrane's core. Here, polymer translocation can be considered as the diffusion of its center of mass along an effective free energy landscape determined by the self-assembled membrane environment~\cite{katzMethod1974,diamondInterpretation1974,sommerCritical2012}. Translocation of homopolymers through bilayer membranes was recently understood theoretically by means of propagators as the solution of Edwards equation \cite{wernerThermal2017} in good agreement with coarse grained simulations~\cite{wernerHomopolymers2012}. Any inhomogeneity in polymer sequence renders the problem more complex being then equivalent to solving Schroedinger's equation in time-dependent potentials. Random sequences can only be treated analytically when assuming a well-defined distribution of segment properties, and predictions such as the adsorption transition at hard walls consider the limit of infinite chain length~\cite{soterosStatistical2004}. Further complexity is expected when attempting to predict theoretically the translocation time of cell-penetrating-peptides through a soft boundary (membrane) while having particular bending stiffness and dihedral potentials given by the amino acids in the sequence. Coarse grained simulation results on random copolymers indicate that the main factors for copolymer translocation are their average hydrophobicity as well as their degree of adsorption at the membrane-solvent interfaces~\cite{wernerTranslocation2015}, which shall be reflected in the main modes of their potential of mean force. A rigorous theoretical description as a function of sequence, however, is missing to date. The lack of theory does meanwhile not exclude the recent progress in finding artificial cell-penetrating peptides and antimicrobial peptides by brute-force parallel screening~\cite{marksSpontaneous2011,kauffmanMechanism2015,fuselierSpontaneous2017,kauffmanSynthetic2018} that may even outperform evolutionary highly conserved Tat- or penetratin-based sequences for biomedical application. Wimley et al. found that fine-tuned differences in short-block amphiphilic have significant effect on peptide translocation rates following rules that seems not obvious at the moment. In turn, sequences leading to optimal points in their biomedical performance can possibly be found in unexpected corners of sequence space, that are potentially accessed by sequence-cargo-co-evolution~\cite{kauffmanSynthetic2018}. 

For artificial neural networks, the database for training is the crucial factor when determining their generalization performance. For soft matter objects it is very costly to obtain a reasonable training sets via experiments or computer experiments such as molecular dynamics, dynamic Monte Carlo, or even mean-field methods, since a single training data point requires typically minutes to several CPU days. In this work, we will have the luxury of having a complete sequence-to-property map availlable for training and testing NN algorithms thanks to the GPU-accelerated sampling of random polymer configurations for a given sequence. GPU-accelerated Rosenbluth-Rosenbluth sampling of an amphiphilic copolymer brought into a model representation of lipid membranes allowed us to generate a significant number of configurations for all possible binary A-B sequences for chain length up to $N=16$. Based on this data, a neural network is trained in order to predict mean first escape times of the polymer through the layer. Interestingly, by restricting the training set to a narrow window of translocation times, where the largest and slowest are separated by a factor of 30 only, the network is capable of predicting the correct order of magnitude of translocation time for all other sequences although they are spread by more than 9 orders of magnitude in absolute value.

The rest of the paper is structured as follows: In section \ref{sec:method}, we describe the computational methods and the chosen example of polymer translocation through a double-layer interface as function of a binary A-B- amphiphilic copolymer sequence. 
In section \ref{sec:RSres} we introduce the results of the Rosenbluth-Rosenbluth for translocation time prediction, and by comparison with free energy estimates of self-avoiding walks near interfaces we underline the physical meaning and richness of the results. In section \ref{sec:NNres} we describe the results of neural network based translocation time prediction based on two different training schemes. In section \ref{sec:conclusions} we summarize the results.



\section{Methods}\label{sec:method}

\subsection{\label{sec:RS}Rosenbluth-Rosenbluth Sampling (RS)}

\begin{figure}[t]
	\centering
	\includegraphics[width=8.0cm]{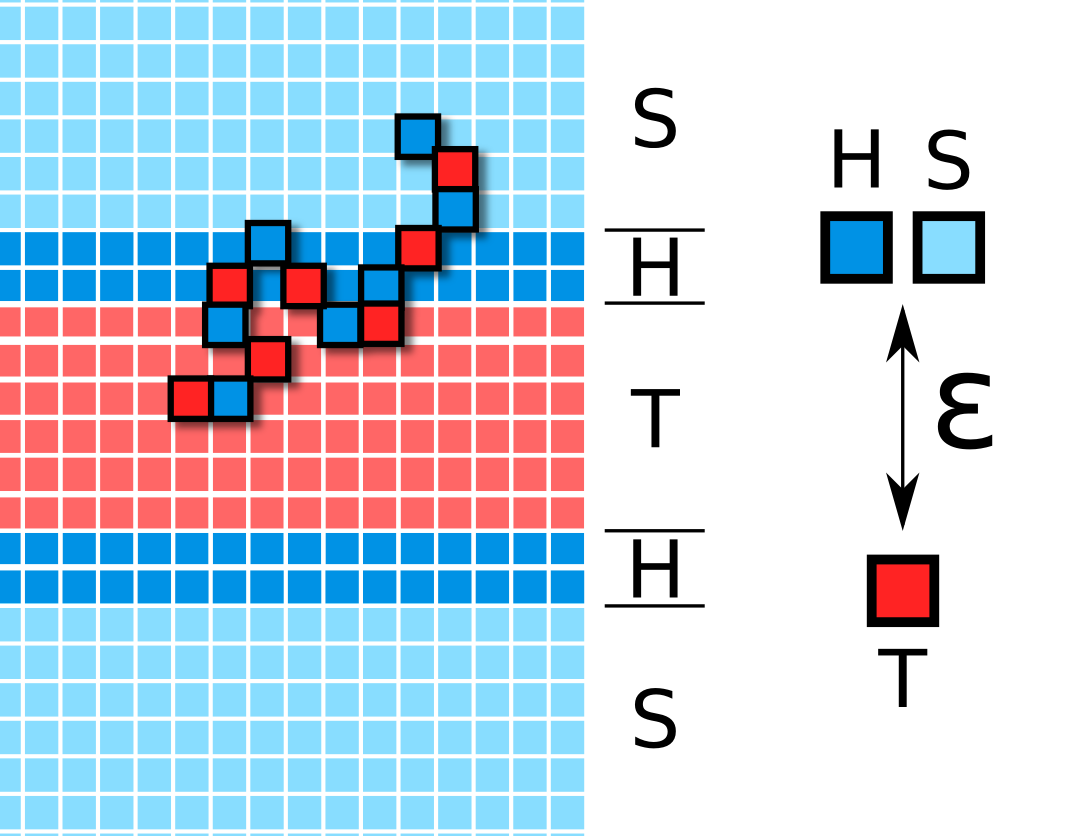}
	\caption{Illustration of a coarse grained polymer chains as used in our simulations within the grid occupancy of a laterally homogeneous membrane, and repulsive interactions between effectively two components (H,S) and (T).}
	\label{f1}
\end{figure}

We consider the diffusive transport of a polymer through a lipid membrane resembling a homogeneous oil slab as shown in Fig. \ref{f1}. In particular, we are interested in mean first escape time a polymer through the membrane as a function of length, $N$, sequence of hydrophilic head (H) and hydrophobic tail monomers (T). Coarse grained polymers are embedded into an external concentration field that represents bilayer membrane on a mean-field level composed of an hydrophilic region (H), and a hydrophobic core (T), as well as solvent (S). The hydrophobic core has a thickness of 6 lattice units.

Monomers are represented as single cell occupations on a simple cubic lattice, and bond vectors are taken from a set of 26 vectors with lengths of $1$, $\sqrt{2}$, and $\sqrt{3}$ lattice units. Double occupancy of lattice sites is forbidden, and the monomers have excluded volume. This set of static rules corresponds to those of Shaffer's Bond Fluctuation Model~\cite{shafferEffects1994}.

Between hydrophilic sites (H and S), and hydrophobic sites (T), we implement short-range repulsive interactions.
We write the internal energies of H and T monomers of the polymer as
\begin{equation}
\begin{aligned}
U_H(\vec r)  = \epsilon c_T(\vec r);~~U_T(\vec r)=\epsilon ( c_S(\vec r) + c_H(\vec r) )
\label{eq:U}
\end{aligned}
\end{equation}
where $c_x(\vec r)$ are the number of lattice occupancies by species $x$ on the 26 nearest neighbor sites\cite{doteraDiagonal1996}. In order to keep the model simple, we use only a single interaction parameter defined as $\epsilon=0.1k_BT$ with $k_B$ being Boltzmann's constant, and $T$ the absolute temperature. For the enumeration of $c_x $, both the occupancy of the lattice by a given external concentration field (Fig. \ref{f1}) as well as monomer-monomer interactions are taken into account in a way that contacts with the external field are screened by surrounding monomers. Thereby solvent-induced effects on polymer conformations are well represented by the model.

For a given amphiphilic sequence, we aim to calculate the mean first escape time of a polymer between a repulsive boundary at z=-a and an absorbing boundary at z=+a, (Fig. \ref{f1}),
\begin{equation}
\begin{aligned}
\tau_{te}= \frac{1}{D}\int _{-a}^{+a}\mathrm{d}zp^{-1}(z)\int _{-a}^{z}\mathrm{d}z'p(z')
\label{eq:mfet}
\end{aligned}
\end{equation}
where $D$ is the diffusion constant of the polymer, and $p(z)$ is the probability distribution to find the center of mass of the polymer at a given distance, $z$, from the bilayer's mid-plane. The probability distribution $p(z)$ is calculated by generating $M$ polymer conformations $\vec R = (\vec r_1, \vec r_2, \dots, \vec r_N)$ according to the Rosenbluth-Rosenbluth (RS) scheme~\cite{rosenbluthMonte1955}. For each conformation $\vec R$, the contact energy $U(\vec R)$ is calculated according to $U(\vec R) = \sum_{i=1}^N U_{X}(\vec r_i)$ in units of $k_B T$ according to Eq.~(\ref{eq:U}) depending on the species $X$ of the monomer $X=H$ or $X=T$. The center of mass $\bar z(\vec R)=(1/N)\sum_{i=1}^N \vec r_i \vec e_z$ evaluated with $\vec e_z$ being the lattice unit vector along the membrane's normal direction. The distribution $p(z)$ is then written as
\begin{equation}
\begin{aligned}
p(z) = \frac{1}{M}\sum_{i, \bar z(\vec R_i) \simeq z }^{M} w_i \mathrm{e}^{-\beta U(\vec R_i)}
\label{eq:pz}
\end{aligned}
\end{equation}
where the condition below the sum illustrates that only those conformations contribute whose center of mass, is found within a grid distance $(z-1/2) < \bar z \leq (z+1/2)$ from $z$, and $\beta\equiv1/(k_BT)$. In eq~(\ref{eq:pz}), $w_i$ is the Rosenbluth weight of the $i$-th conformation.

For a given sequence of H and T monomers in a polymer backbone, we calculate the mean first escape time according to \ref{eq:mfet} based on the generation of $M = 1.5\times10^7$ RS-generated chains at uniformly distributed random positions within a periodic lattice of $64\times64\times64$ lattice sites. The algorithm is implemented for graphics processing units\cite{guoGPU2017}. In order to analyze how the mean first escape time depends on the amphiphilic sequence of the polymer, we perform the procedure for all $2^N$ H/T sequences for various degrees of polymerization $N \leq 16$.

\subsection{\label{sec:NN}Multi Layer Artificial Neural Network (NN)}

\begin{figure}
	\centering
	\includegraphics[width=\columnwidth]{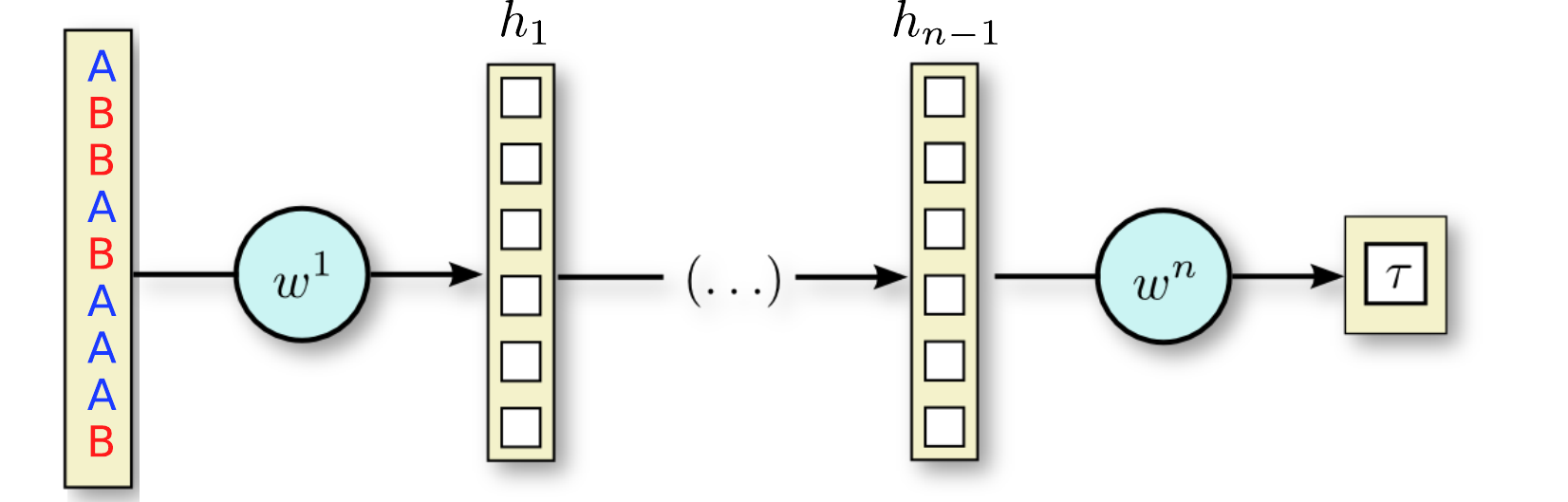}
	\caption{Neural network architecture for translocation time prediction of a polymer as a function of A/B sequence.}
	\label{f5}
\end{figure}

We employ a fully connected neural network involving $\tanh()$-activation as sketched in Fig.~\ref{f5}. The network is composed by two hidden layers with 64 nodes each followed by two hidden layers with 32 nodes each. The input layer corresponds to a vector of values 0 and 1 representing the considered amphiphilic sequence of hydrophobic (0) and hydrophilic (1) monomers. The output layer consists of only one neuron that is compared to the RS-based $\tau$ value for this sequence. The total network depth is $n=5$. Since absolute values of $\tau$ spread over several orders of magnitude, we perform the training with respect to its logarithm. The RS-based values of $\log(\tau)$ are further linearly normalized and centralized into an interval $[-0.9, 0.9]$ in order to be conveniently expressible by the $\tanh$ activation output.

All weights are initialized with uniform random numbers in an interval $[-0.35, 0.35]$.
The feed-forward (ff) back-propagation (bp)~\cite{rumelhartLearning1986} algorithm is employed for training. Error bp is performed after each ff cycle for a randomly selected sequence taken from the training set. The squared difference between the resulting activation at the output neuron and the RS-based $\tau$ value is propagated back as the error for weight adjustment. We set the initial training rate to $\eta=0.02$, which gets reduced by a factor of $(1/1.3)$ every $10^3$ epochs in order to avoid frustration- or early over-training effects. One epoch is defined as the average number of ff-bp cycles per sequence-$\tau$ pair. We set the total number of epochs to $10^4$.

For each sequence we define an unique integer identifier, $1\leq id\leq \ss$, that is sorted according to the RS-based $\tau$-value. A lower $id$ means a lower $\tau$. The whole of $\ss$ sequences is divided into a training set of size $\ss_{train}$, and a test set of the size $\ss_{test}=\ss-\ss_{train}$. For the test set, we define a unique identifier $id_{test}$ for each sequence that is the analog to $id$ for the total sequence space. The index $id_{test}$ labels sequences that are totally unseen by the network during training.

\section{\label{sec:RSres}Rosenbluth-Rosenbluth-Sampling Results}


\begin{figure}
	\centering
	\includegraphics[width=\columnwidth]{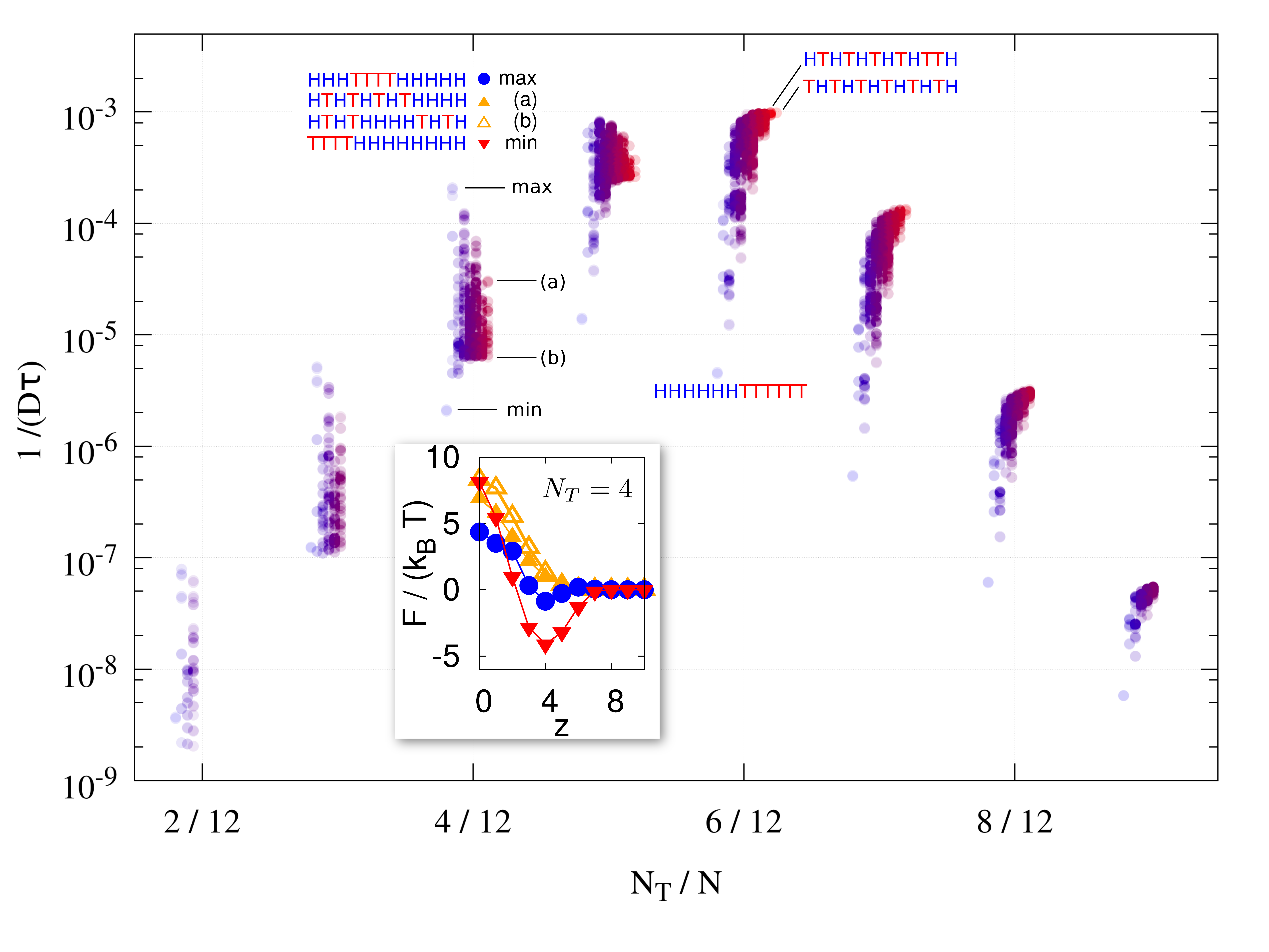}
	\caption{Inverse mean first escape times as a function of the fraction of hydrophobic monomers of a polymer of length $N=12$. Results are shown for all sequences containing between 2 and 9 T-type monomers. Results sharing the same number of T-type monomers are spread within windows of width 0.04 along the ordinate according to the number, $n_b$, of H and T blocks within the sequence. The exact position along the ordinate is calculated as $N_T / N + 0.04\times[n_b/N - 1/2]$. Results for eight sequences are highlighted by labels.}
	\label{f2}
\end{figure}

\begin{figure}
	\centering
	\includegraphics[width=\columnwidth]{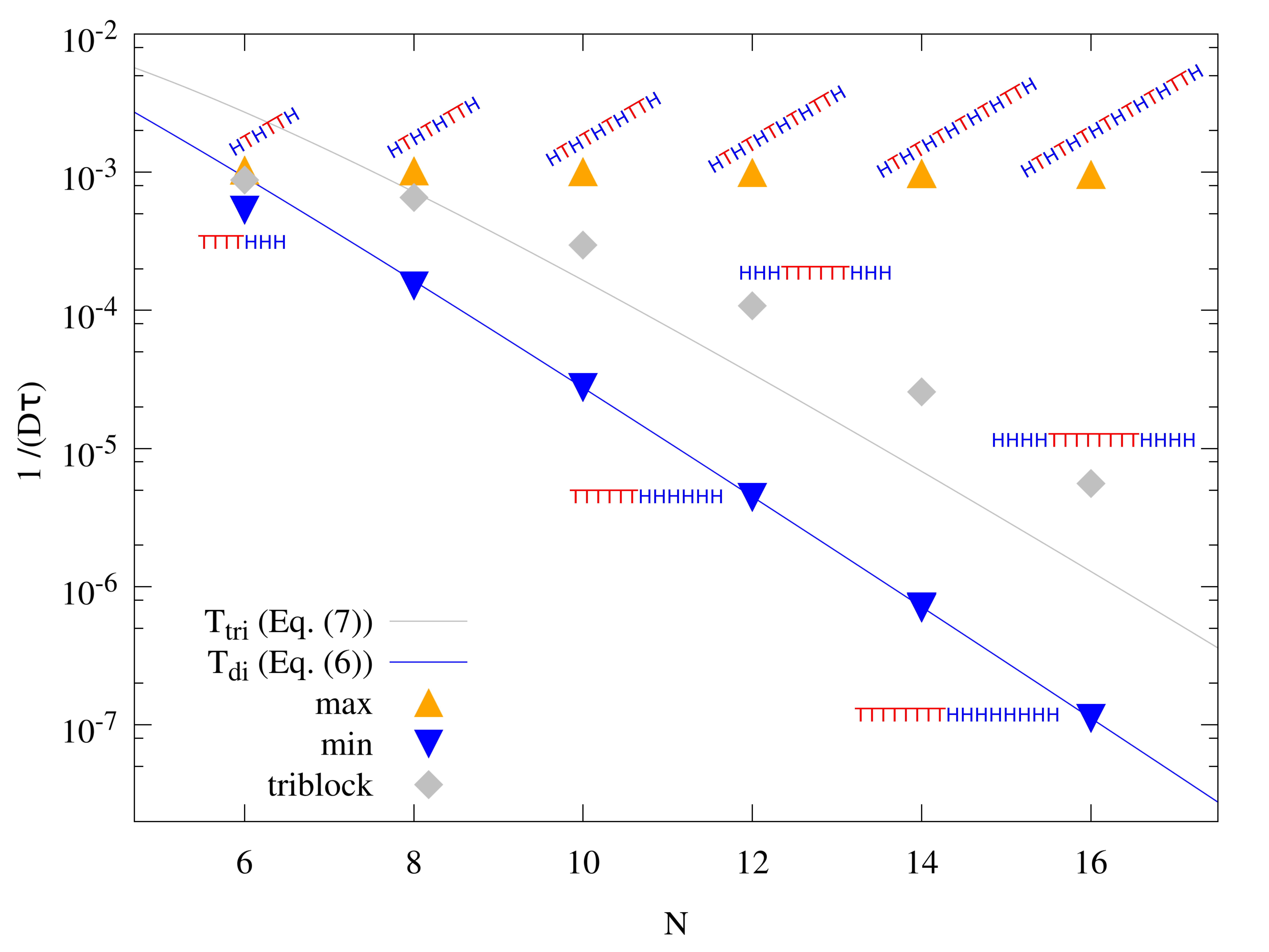}
	\caption{Inverse mean first escape times as a function of the chain length for fractions of hydrophobic monomers of $1/2$. We show results for sequences leading to maximal (minimal) inverse escape times.}
	\label{f3}
\end{figure}

Let us consider the inverse mean first escape time $1/\tau$ as a measure for the frequency of translocation of a polymer through the membrane, which is presented in Figs.~\ref{f2} as a function of the mean hydrophobic fraction along a backbone of $N=12$ monomers. Results for all sequences are shown, and grouped into point clouds centered at the corresponding ratios $N_T / N$. The point clouds are shaped according the number, $n_b$, of blocks of H and T species along the sequence in a way that the points on the right hand side of a cloud represent a polymer with a larger number of blocks.

The results in Fig.~\ref{f2} confirm earlier predictions~\cite{wernerHomopolymers2012} that a maximum of translocation frequency is found near a point of balanced hydrophobicity of the polymer as given by a balanced fraction of H and T units $N_T / N \sim 1/2$, in case that the typical block-size is in the order of the Kuhn-segment of the polymer~\cite{wernerTranslocation2015}. 

In Figure~\ref{f3}, we show the monomer sequences leading to largest and lowest translocation frequency $1/\tau$ as well as the results for triblock copolymers as a function of chain length for the balanced ratio $N_T/N=1/2$ of hydrophobic beads. Results are shown in re-scaled form compensating the chain-length dependence of the diffusion constant, $D$. For alternating sequences, the re-scaled translocation frequency remains in the same order of magnitude showing that the polymers are below the adsorption threshold for the given chain lengths. For polymers that are significantly localized at the membrane-solvent interface one would expect that the desorption to be the rate limiting process for translocation. Adsorption effect are clearly visible for diblock copolymers showing a nearly exponential decay of translocation frequency as a function of chain length. For diblock copolymers we expect that the desorption of the hydrophobic block from the membrane is the most significant rate-limiting process, and consequently diblock sequences lead to minimal translocation frequencies. It is important to notice that for hydrophilic blocks larger than the membrane width, the switch of an hydrophilic end from one solvent side to the opposing solvent does only require a limited number of hydrophilic beads to be in contact with the lipid core at the same time, whereas the escape of the hydrophobic block into the solvent requires all monomers of the block to be displaced into solvent environment. Dynamic barriers such as the steric hindrance of the polymer backbone by lipid tails, is, however, not included in the mean-field environment.

In Fig.~\ref{f3} it becomes visible that, the symmetry of the polymer sequence with respect to hydrophilic ends adds an important factor to the desorption probability. In particular, when comparing results for triblock copolymers where the longest chains show a more than one decade larger translocation frequency as compared to diblocks. The difference can be understood qualitatively by estimating the adsorption free energy in the strong segregation limit as
\begin{equation}
\begin{aligned}
\Delta F_{ads}(N) = -C \epsilon_0 N_T + \Delta F_{el}
\label{eq:Fads}
\end{aligned}
\end{equation}
where $C$ is the average number of contacts of $T$-monomers with the lipid environment (coordination number), and $\Delta F_{el} = -k_B T \ln[Z_{surf}/Z_{free}] $ is an elastic contribution due to the reduction of the partition function from $Z_{free}$ to $Z_{surf}$ upon localization at the surface. The partition sum for a self-avoiding walk takes the form~\cite{grassbergerMonte1993,duplantierPolymer1986,degennesScaling1979}
\begin{equation}
\begin{aligned}
Z(N,\epsilon_{self}) = q(\epsilon_{self})\mu^N N^{\gamma - 1}
\label{eq:Zel}
\end{aligned}
\end{equation}
where $q$ is a non-universal function on the particular form of short-range interactions ($\epsilon_{self}$), and $\mu$ is the critical fugacity for the  given random walk logic and lattice. The exponent $\gamma$ depends on the topology of the polymer that is either in free solution or attached to a surface. One applies $\gamma \equiv \gamma_1 \approx 0.678$~\cite{heggerChain1994,grassbergerSimulations2005,clisbyThreedimensional2016} for strands having one end grafted, and $\gamma \equiv \gamma_{11} \approx -0.39$~\cite{grassbergerSimulations2005,clisbyThreedimensional2016} for strands having both ends surface-attached. The partition sum in free solution scales as $Z_{free}\sim \mu^N N^{\gamma_0-1} $ with $\gamma_0\approx 1.1567$~\cite{hsuScaling2004,schramExact2011,clisbySelfavoiding2007}. Since we further compare only ratios of partition sums for given total chain length, we assume that $q$- and $\mu$-dependent contributions cancel up to a factor of the order unity.

\begin{figure}
	\centering
	\includegraphics[width=\columnwidth]{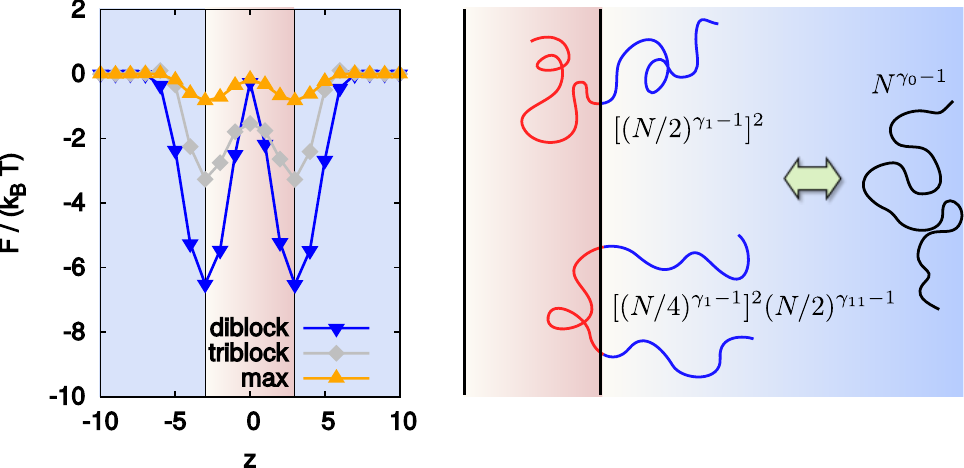}
	\caption{Free energy profiles for various polymer architectures ($N=12, N_T=6$), and corresponding relevant states for estimating desorption probabilities.}
	\label{f4}
\end{figure}

The probability density to find a symmetric diblock copolymer in bulk solvent as compared to a state adsorbed at an interface as illustrated in Fig.~\ref{f4} then reads
\[
 p_{di}(N) = \exp(-\beta(F_{free} - F_{ads})) = \mathrm{e}^{-c \epsilon N/2}\frac{ N^{\gamma_0-1} }{(N/2)^{2(\gamma_1-1)}}
\]
Now, assuming that the desorption is the rate limiting process, we write the estimate for the translocation frequency as
\begin{equation}
\begin{aligned}
T_{di} = T_0 p_{di}
\label{eq:T}
\end{aligned}
\end{equation}
In Fig.~\ref{f3} we show the results for Eq.~(\ref{eq:T}), where $T_0=0.123$ and $C=19.6$ have been adjusted for obtaining least-squared differences from the diblock RS-results. The results confirm the dominance of the exponential factor dominated by pair-interactions of the hydrophobic block.

The ratio between partition sums for interface-adsorbed diblocks and triblocks allows to project from diblock to triblock predictions for translocation frequencies,
\begin{equation}
\begin{aligned}
T_{tri} =  2^{2(\gamma_1-1)}\left(\frac{N}{2}\right)^{\gamma_{11}-1} T_{di}
\label{eq:rGrass}
\end{aligned}
\end{equation}
which is plotted in Fig.~\ref{f3} for comparison. The resulting up-shift catches up to the RS-diblock results up to a factor corresponding to a remaining free energy difference of $1.4\mathrm{k_BT}$ that is missing in Eq.~(\ref{eq:rGrass}).

With this discussion in mind, it is interesting to have a look back to Fig.~\ref{f2} for understanding surprising features observed in the sequence maps of slightly hydrophilic polymers. By the example of a fraction of $4/12$ of hydrophobic monomers, we demonstrate that the polymers comprising the shortest amphiphilic blocks (``(a)'' and ``(b)'') are found in a middle range of translocation frequencies, while triblock copolymers similar as those discusses in Figs.~\ref{f3} and \ref{f4} lead to the largest translocation frequencies. A comparison of the free energy profiles shown as an inset in Fig.~\ref{f2}, underlines the interplay between surface adsorption and hydrophobic / hydrophilic balance that leads to the result. Short-block (``(a)'' and ``(b)'') are mainly subject to an effective free energy barrier for insertion into the bilayer's that is the rate-limiting factor for translocation. The result reflects the fact that the polymer is effectively hydrophilic, and shows negligible surface adsorption effects. Combining T-monomers into a larger center block, however, allows for anchoring of the polymer at bilayer-solvent interfaces, and thereby effectively reduces the rate-limiting repulsion from the membrane environment. On the other hand, for the diblock copolymers with $N_T/N=4/12$, adsorption at the bilayer-solvent interface turn over to dominate the free-energy profiles, and lead to the largest escape times found for the given hydrophilic / hydrophobic ratio. 

From the comparison between RS-results for $\tau$ and previous literature we therefore conclude that the dynamic interpretation of the sampling results is reasonable.

\section{\label{sec:NNres}Machine-Learned Sequence to Translocation Mapping}

The massive data sets generated by GPU-accelerated RS sampling form a powerful basis for the machine-learning based search for sequences fulfilling given criteria. Although a network similar to Fig.~\ref{f5} can be designed in order to predict a general functional of sequence $\vec Y(\vec X)$, in this work let us stick to the example of translocation times, $\tau$. In the following, we pick the example of a chain length of $N=14$ monomers. The total number of sequences excluding the mirror-symmetric ones is $\ss=8256$. The fraction of sequences within the training set we fix to $f_{train}\equiv\ss_{train}/\ss = 1/7$. The ratio between training to the remaining test set results in $1:6$. However, we follow two distinct schemes for the distribution of training sequences within the sequence space: In the $uniform$ scheme, we define equidistant intervals of size $1/f_{train}$, along the $\tau$-sorted sequences ($id$-space) and select the central sequences within each interval as the training set. In contrast, in the $\tau-window$ scheme we select every second sequence within a window $\ss/2 < id \leq \ss/2+2\ss_{train}$. Note that thereby we select sequences within a narrow window in the upper half of translocation times.

\begin{figure}
	\centering
	\includegraphics[width=\columnwidth]{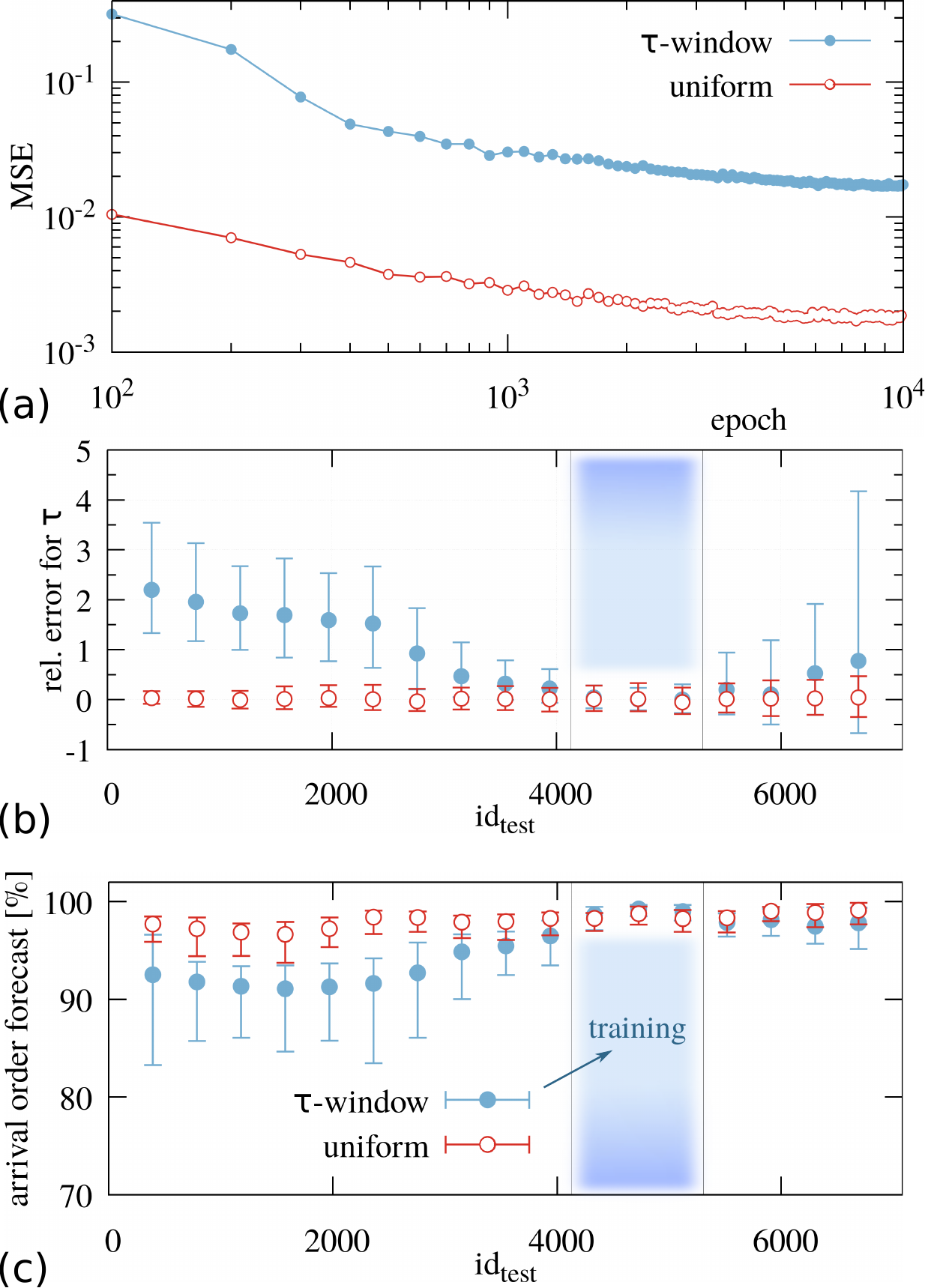}
	\caption{(a) Evolution of the mean squared error (MSE) for the test data set as a function of training epoch. (b) Relative error of the predicted value of $\tau$ according to eq.~(\ref{eq:dtau}) as a function of sequence index $id_{test}$ in the test set. The result is averaged for 17 groups (bins) of sequences along the RS-$\tau$ sorted test set ($id_{test}$). Error bars denote a confidence interval of $90\%$. The blue box labels the range of sequence of the test set in case of $\tau-window$. (c) Percent of correctly predicted faster or slower other sequences as a function of sequence index. Here we use the same binning and error bar definition as in (b).}
	\label{f6}
\end{figure}

\begin{figure*}
\includegraphics[width=\textwidth]{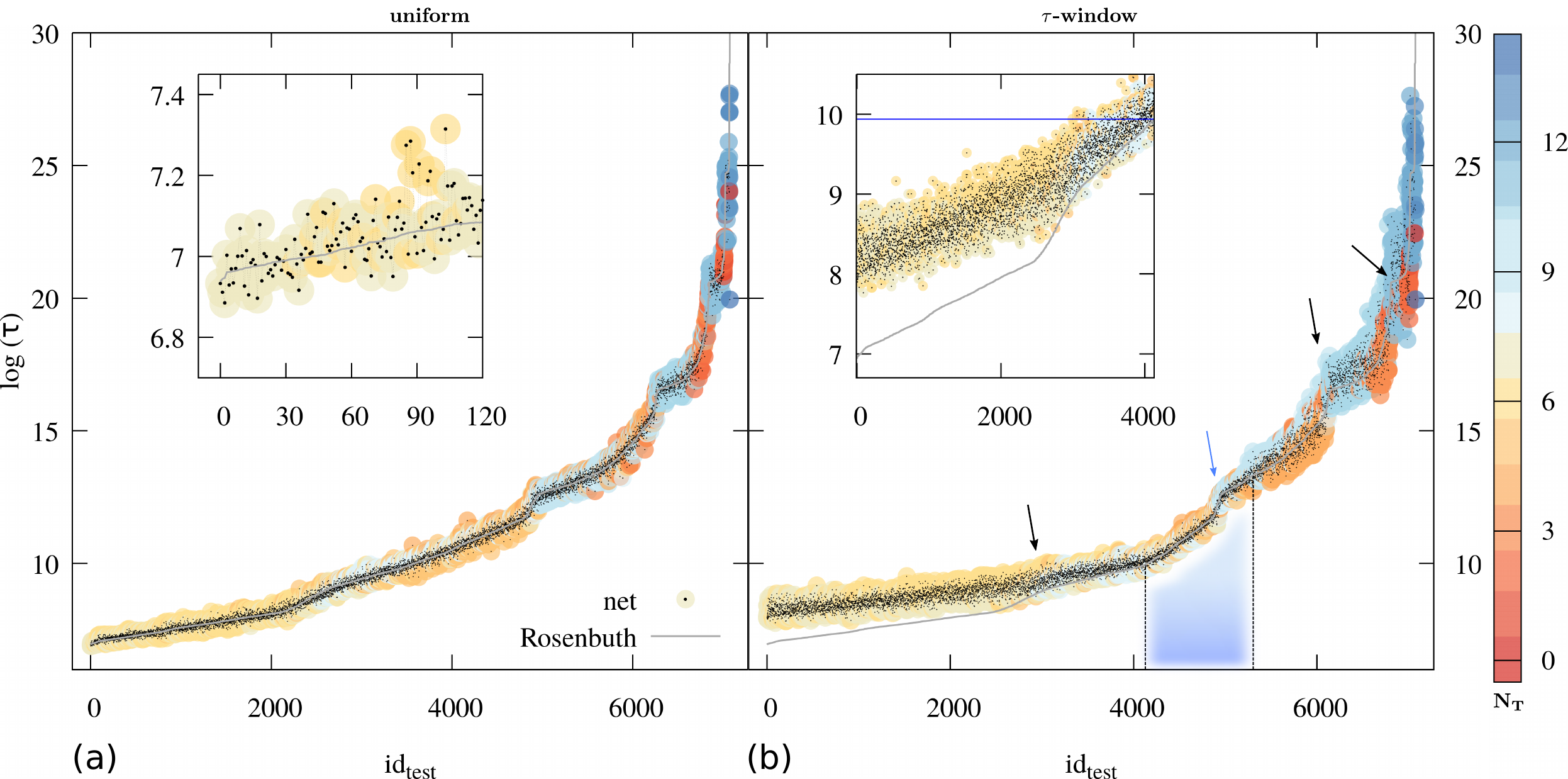}
\caption{Neural network prediction (dots) for the mean first escape time $\tau$ for unseen data (test set) are compared to RS-based results (grey line). Data is shown as a function of a unique identifier $id_{test}$ for sequences in the test set, that is sorted according to the RS-bases result for $\tau$. The ratio between training- and test set sizes is $1:6$. The number of hydrophobic units, $N_T$, is shown as color-coded halos. In (a) the $uniform$ training set distributed homogeneously along the full RS $\tau$-sorted sequence list. In (b) we chose every second sequence within the blue labeled $\tau-windows$ range between $id_{test}=4128$ and $id_{test}=5305$. Test set sequences are skipped in this plot such that the slope is doubled as compared to the full sequence set within the labeled interval. The insets show details in the fields of lowest $\tau$. The blue horizontal line in (b) inset labels the RS-based $\tau$-value at the lower bound of the $\tau-window$.}
\label{f7}
\end{figure*}

In the Figures~\ref{f6} and \ref{f7} we summarize the results of the training, and the performance of the resulting network with respect to the test set. 

In Figure \ref{f6}(a), the development of the mean squared error (MSE) between RS-based 
$\log(\tau)$ values and the output neuron activation for all test sequences (unseen) is presented. We note a reliable convergence of MSE values for both $uniform$ and $\tau-window$ training sets towards a horizontal line indicating that training was stopped early enough for not running into over-training. In case of the $uniform$ training set, MSE results typically end up at one order of magnitude lower as compared to the $\tau-window$ training set. The corresponding root mean squared deviation from the expected value typically reduces by a factor of $\sqrt 10\sim3$. For the $uniform$ training set, the root MSE (see Fig.~\ref{f6}) points to a typical error of the is $\log(\tau)$ prediction of $\sqrt(\mathrm{MSE})\sim4.5\%$, whereas for the $\tau-window$ training set we observe values of $\sqrt(\mathrm{MSE})\sim14.1\%$. 

In Figure \ref{f6}(b) we show the corresponding mean relative error for the back-converted (not logarithmic) time $\tau$ according to
\begin{equation}
\begin{aligned}
\frac{\Delta \tau}{\tau} = \exp[\Delta \log(\tau)]-1
\label{eq:dtau}
\end{aligned}
\end{equation}
where $\Delta \log(\tau)$ is the absolute difference between the RS-based $\log(\tau)$-value and the output neuron activation. For the $uniform$ training set, the relative error scatters symmetrically between $-36\%$ and $+47\%$ as found for largest index $id_{test}$ (largest $\tau$), whereas for the fastest polymers $90\%$ of sequences stay within a error of $-9\%$ to $+17\%$. For this training set, equivalent to a random selection of sequences, such high accuracy of the network prediction is remarkable when seeing that the ground truth in form of RS-based values of $\tau$ is spread by a maximum factor of $\tau_{max} / \tau_{min} \sim 2\times 10^{11}$. For the $\tau-window$ training set, the relative error far away from the training window increases as compared to the $uniform$ set. Nevertheless, as the maximum range of relative errors is found in the interval of $-0.67\leq\Delta\tau/\tau\leq4.17$ for the largest index $id_{test}$, we conclude that typically the prediction hits the right order of magnitude for $\tau$ despite the fact that we used only the narrow sequence window for training. It is interesting to note that the translocation times of the fastest sequences is typically predicted correctly by a factor of $\sim 3$ despite the large distance from the training window.

Absolute values are not always the main question for the modeled mapping $\vec Y (\vec X)$, and in some cases it is enough to obtain a decision statement upon the performance of two structures. When comparing two polymer sequences, for instance, we may ask which of those translocates faster. In Figure \ref{f6}(c) we therefore show the performance of the trained network to give the right answer for this question as a function of sequence $id_{test}$. On average, in case of $uniform$ training $98.1^{+0.8}_{-1.7}\%$ of other sequences are correctly attributed as slower or faster (with a confidence of $90\%$), and for the $\tau-window$ training set $95.1^{+1.8}_{-4.1}\%$ of pairs are correctly labeled. For the $\tau-window$ training set, the performance far away from the training window is reduced in particular for sequences with a lower $id_{test}$ index. However, the average fraction of correct decisions does not drop below $93.3\%$ for the selected bin size.

By Figure \ref{f6} we therefore demonstrated that a quantitative prediction of translocation times is possible by the applied ML model, and the accuracy depends crucially on the distribution of training sequences.

In Figure \ref{f7} we outline more details of the training result by showing the predicted value of $\log(\tau)$ for the whole test sets $uniform$ and $\tau-window$ in Fig. \ref{f7}(a) and Fig. \ref{f7}(a), respectively. The monotony of predicted data points for both training sets follows the base data line despite the scattering of the data as discussed for Fig. \ref{f6}. In particular, for the $\tau-window$ training set, we emphasize that the order of translocation times is predicted correctly for the fastest sequences although the training set covers only a narrow window within the slower half of sequence. The statistical scattering of prediction is likely to be reducible via training a number of networks with independently seeded weights, and averaging the prediction from the ensemble of networks. Note that the training typically required only several minutes on a single CPU thread.

Another interesting observation is the prediction of step-like features in translocation time (arrows in Fig. \ref{f7}(b)) as function of $id_{test}$, that are reproduced throughout the test set although located outside of the $\tau-window$ training range. Thus, even the relatively simple network seems capable of finding a generic rule that links sequence and translocation time, and thereby expresses the rather rich result based on Equations (\ref{eq:mfet}) and (\ref{eq:pz}) without knowledge of confirmation entropy nor the Kramer's integral. In view of the generalization performance observed for the $\tau-window$ training set it therefore seems that the network developed an implicit internal representation approximating the mathematical rules linking copolymer sequence and translocation that involve the partitioning of self-avoiding walks in external fields, and integral Equation (\ref{eq:mfet}).


\section{\label{sec:conclusions}Conclusions}

We apply a massively parallel sampling of the conformations of amphiphilic copolymers by means of self-avoiding random walks within a given density field representing a model for amphiphilic bilayer membranes. We estimated the free energy profiles of the polymers composed of hydrophilic (H) and hydrophobic beads (T) with respect to distance from the membrane as a reaction coordinate. We calculated the mean first escape time $\tau$ as a measure for polymer translocation time through the model membrane all $2^N$ binary sequences up to chain length $N\leq16$. Our results confirm that polymer translocation is controlled by a balance of the overall hydrophobicity of the polymer, and is inhibited by adsorption at the bilayer-solvent interfaces~s\cite{sommerCritical2012,wernerHomopolymers2012,wernerThermal2017,wernerTranslocation2015}, which is consistent with the picture for small solutes \cite{marrinkPermeation1996} and larger solid objects such as carbon-nano-tubes \cite{pogodinCan2010}.

Amphiphilic polymers at a balanced hydrophobicity show smallest translocation times when the sequence exposes small repeating amphiphilic features, while longest waiting times are associated with a diblock structure of the whole chain. The different translocation rates between diblock and triblock copolymers as well as their chain-length dependence can be explained qualitatively when comparing adsorption free energies at the bilayer-solvent interface involving surface-critical exponents. The relatively weak dependence of the translocation time of balanced hydrophobicity small-block alternating copolymers from chain length indicates that local amphiphilic features are only weakly interacting with the bilayer-solvent interfaces and the copolymer polymer effectively resembles a homopolymer chain for which the membrane is energetically transparent. Chain-length dependence in this case is expected to increase when effective monomer association constants are stronger than in the present model. When considering slightly hydrophilic backbones larger hydrophobic start to become more prominent in sequences leading to smallest translocation times as they promote the association of the net-repulsive backbone from with the hydrophobic membrane core.

The extensive data base generated by RS-sampling has been used to feed a multi-layer artificial neural network (NN) machine learning algorithm with four hidden layers in order to explore the capability of so-called deep learning approaches for finding a general rule of how copolymer sequence translates into translocation times through biological barriers. We demonstrate that even by using a low fraction $1/7$ of randomly selected training examples as compared to the total number $2^N$ of binary sequences for $N=14$, the NN achieves a root mean squared deviation in the order of $4.5\%$ for the logarithmic mean first escape time $\log(\tau)$. In order to test the generalization performance of the network, we performed a second training process, where training examples have been selected from a narrow window of sequences with respect to translocation times $\tau$ covering a factor of $\approx 30$ between maximum and minimum translocation times contained in the training set. In this case, the network predicted correctly the order of magnitude of the training set covering a much wider range between maximum and minimum value of $\tau$ separated by $~10$ orders of magnitude. We conclude that the multi-layer perceptron developed an internal representation of the mathematical rules linking sequence and translocation times. The network thereby encodes a complex interplay between polymer net hydrophobicity and sequence-dependent adsorption at the bilayer-solvent interfaces that to date can be treated in theoretically closed form only for simple boundary cases as it involves the sequence-dependent polymer confirmation entropy, and solving the diffusion problem in inhomogeneous free energy-landscapes.


%

\end{document}